# Transcribing medieval manuscripts for machine learning

**Estelle Guéville[1], David Joseph Wrisley[2] ***

1 Yale University, United States of America

2 New York University Abu Dhabi, United Arab Emirates

*Corresponding author: David Joseph Wrisley: djw12@nyu.edu

**Abstract**

This article focuses on the transcription of medieval manuscripts. Whereas problems of transcription have long interested medievalists, few workable options in the era of printed editions were available besides normalisation. The automation of this process, known as handwritten text recognition (HTR), has made new kinds of digital text creation possible, but also has foregrounded the necessity of theorising transcription in our scholarly practices. We reflect here on different notions of transcription against the backdrop of changing text technologies. Moreover, drawing on our own research on medieval Latin Bibles, we present general guidelines for customizing transcription schemes, arguing that they must be designed with specific research questions and scholarly end use in mind. Since we are particularly interested in the scribal contribution to the production of codices, our transcription guidelines aim to capture abbreviations and orthographic variation between different textual witnesses for downstream machine learning tasks. In the final section of the article, we discuss a few examples of how the HTR-created transcriptions allow us to address new questions at scale in medieval manuscripts, such as textual variance across witnesses, the prediction of a change in scribal hands within a single manuscript as well as the profiling of individual and regional scribal characteristics.

## I    INTRODUCTION

In the early twentieth century, many scholars focused on the preparation of editions and translations of texts previously available to the few specialists able to read archaic hands and privileged enough to travel to work in person with them. Valuable scholarship in its own right, the preparation of these editions and translations of particular texts were important enough to justify the effort and time, and they laid the foundation for generations of scholarship in medieval studies. On the other hand, for many materials in historical archival collections–including already digitised collections–medievalists have only created partial transcriptions, if any at all. Access to textual material from the medieval period has increased greatly in recent years with digitisation, and we are able to imagine many new lines of research inquiry in decades to come. What challenges do new frontiers of automation in the archives raise with respect to medieval studies and in particular to the ways we transcribe? We already have mature methods for remediating generations of printed editions of medieval works such as Optical Character Recognition (OCR), but we can ask ourselves if these are the kinds of text we want to use for computational analysis with medieval texts in the future. We suggest instead that one way forward is by going "back to the scriptorium," by which we mean articulating systematic transcription guidelines which allow for layers of data to be captured from manuscript witnesses, data which are otherwise lost by normalising methods typical of some forms of editing. If we agree with the notion that normalisation choices are in fact editorial choices, our specific approach is to exercise as few editorial decisions as possible in the process of transcription (Cugliana and Barabucci, 2021). The specificity of transcription guidelines adopted in any given project, we argue, depends on their end use. Whereas in libraries and archives, highly normalising transcription criteria may facilitate keyword spotting and content discoverability for the general reader, philologically minded research may choose more specific criteria which eschew normalisation. Indeed, workflows may indeed emerge in the future which allow for multiple transcriptions to be derived from, and to be linked to, the digitised manuscript folio. Our perspective in this article is that of an expert community of medievalists interested in creating philologically faithful diplomatic transcriptions for the purpose of computational study of scribal profiles. In this article, we argue that if medievalists hope to pursue new kinds of analysis facilitated by advanced computational research, we need to theorise modes of transcription, not only for human reading, but also for machine processing.

## II. PRACTICES OF TRANSCRIBING MEDIEVAL MANUSCRIPTS: A VERY SHORT HISTORIOGRAPHY

In this section, we discuss briefly different ways that editors and publishers of medieval texts have treated the question of the difference between the writing systems that we typically use today and those that are found in manuscripts. This section is not meant to be an exhaustive assessment of historical trends, but rather a convenient way of situating our discussion of transcription. Practices in handwritten text recognition (HTR) are fast moving and expanding to many different fields of manuscript studies and book history. We frame our discussion by referring to work we have done specifically with automatic transcription of thirteenth- and fourteenth-century Latin Bibles and using plain text output, but we trust that our contextualised discussion of transcription will benefit other communities who may be considering automatic forms of text creation for other use cases.

### 2.1 Historicizing Normalisation

The transcription of an old text is both a theoretical and a practical endeavour. We all have inherited multiple methodologies for transcribing, but machine learning systems for handwritten text recognition (HTR) such as Transkribus, eScriptorium and others that will no doubt come into

Journal of Data Mining and Digital Humanities                    http://jdmdh.episciences.org
ISSN 2416-5999, an open-access journal

being in the near future bring to the fore key issues related to working with manuscripts and archival documents, requiring us to think carefully about transcription methods. Whereas historical debates about standards for transcribing documents have typically focused on how best print or digital editions can be made of them, these questions need to be updated, given the emergence of new research infrastructures such as HTR. Three main points related to this necessary update are worth evoking here. First, their emergence emphasises the question of normalisation as a historically contingent and changing category. Second, the rise in popularity of HTR foregrounds the necessity for anticipating how target transcriptions will be used in research in order that the quality of resulting transcriptions matches the way that researchers want to study them.[1] Third, the question arises of how HTR models emerging in the digital GLAM sector–across a spectrum from general to specialised models–for the automatic transcription of text will quickly become part of the research landscape, changing modes of analysis and interpretation in the historical humanities.[2]

So, what are some of the ways that we implicitly or explicitly normalise texts when we transcribe them? Scholars working on a particular source base might have a given set of transcription norms inherited from a publisher or a philological education encouraging the normalisation of letter-forms such as i/j or u/v or imposing specific rules on capitalization or spacing. When looking at the question of transcription, we should not forget that normalisation has itself become a norm; editors of the first editions and incunabula in the fifteenth century often proposed versions of a text much closer to the original manuscript than today's scholars have. As they strove to reproduce medieval manuscripts in a way that they saw them, they maintained many of their features, including not only columns, running titles or rubrics but also special letter-forms and abbreviations, as seen in Figure 1. Special letter types that included these abbreviations were created for incunabula, including what we now call the macron (Unicode 0304) but also several others (ʒ; ⁹; etc.). These incunabula also maintained the distinction between normal and long s (s/ʃ), normal and insular d (d/ð), normal and rotund r (r/ꝛ), a differentiation that was usually collapsed in modern editions of manuscripts. To say the transition to print culture eliminated the need for abbreviations and different letter-forms is false. In later centuries, normalisation also followed the period norms including the use of the long "s" throughout the eighteenth century (Attar, 2010).

---

[1] Little discussion of how we transcribe for HTR is found, perhaps because researchers working with HTR transcription have to a large extent focused on general searchability, rather than text creation for computational study.
[2] A number of different projects around Europe and beyond are investigating the possibilities of HTR from the perspective of multiple languages and textual traditions. A small selection includes Fradejas Rueda (2022), Hodel *et al.* (2021), Camps *et al.* (2019). The question of finding common ground in an ecosystem of proliferating HTR ground truth is addressed by projects such as HTR United (Chagué et al, 2022).

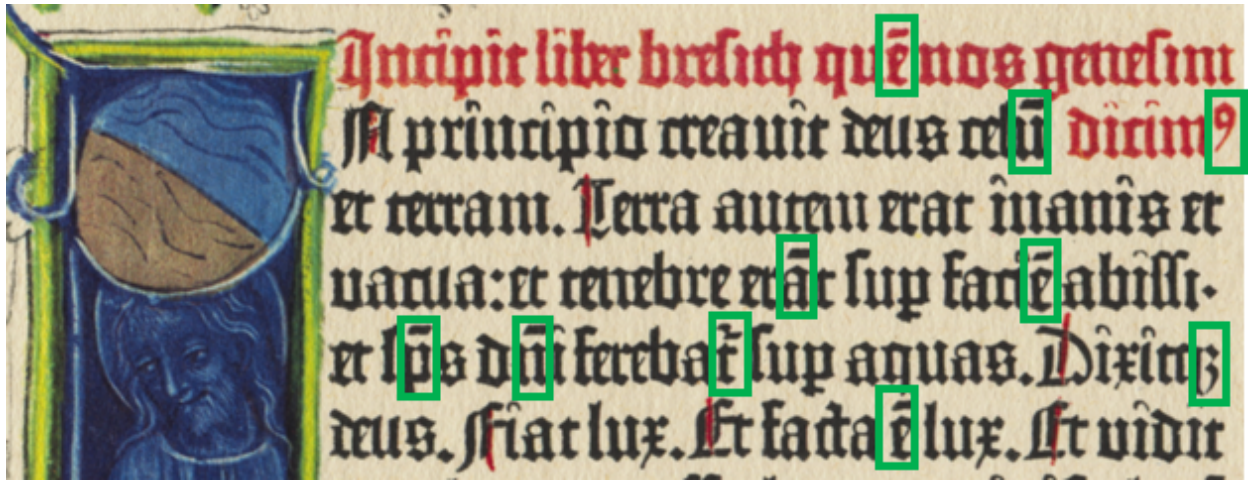

**Figure 1:** An image from an incunabulum Gutenberg Bible illustrating printed versions of special letter-forms and abbreviations (some of which are marked by the green boxes). Source: Beginning of book Genesis in 42-line Gutenberg bible, fol. 5r, vol. 1, Staatsbibliothek Berlin 259, 1454/55.

But what kinds of transcriptions do we find in circulation today? Scholars usually distinguish between normalised transcriptions, semi-diplomatic, and diplomatic ones, although often each transcriber defines their own set of rules, leading to a large variety in transcription norms. In Table 1, we give examples of transcriptions of each type, using a corpus of Latin manuscripts we are studying in the Paris Bible Project.[3] The transcription method illustrated in the first text column (labelled "normalised") changes many letter-forms, capitalisation and spacing, silently expanding abbreviations, correcting the text where the transcriber feels like it is required, and replacing unfamiliar letters with rough equivalents from the Roman alphabet. The text it produces is very easy to read for anyone who is not familiar with palaeography, since it conforms to modern-day literacies and it allows some lines of research inquiry such as literary style, word frequency, or comparisons of texts to be made with limited intervention. At the heart of any form of transcription is, however, the spectre of data loss, that is, how normalisation removes information that is present in a document for its representation as a text.

If, for example, as medievalists, we want to use language resources (LR) and natural language processing (NLP) techniques, especially with text created by the supervised machine learning of HTR, an inevitable conflict between normalised versions of texts and document-level transcriptions lies on the horizon (Piotrowski, 2012; Hodel, 2022). We would like to suggest that the notion of study corpora is intertwined with the idea of normalisation, a set of transcription norms that has been chosen and perfected in the majority of cases since the 19[th] century. In the specific case of the Paris Bible Project our focus has been on specific variation in spelling and abbreviations. Nonetheless, the rise of HTR-created corpora, especially those created with conservative models preserving abbreviations, will no doubt require new methods in NLP to handle non-normalised text with a significant amount of variance.

---

[3] One of the Paris Bible Project's goals is to study micro-features in extant Latin Bibles as a way of performing predictive analysis about copying and scribal habits, as well as reimagining localization or dating. For more information about the project, our working guidelines for transcription, a list of special characters used in transcription and the project blog, see the project site: https://parisbible.github.io/. Our project repositories at GitHub include ground truth samples for reproducing HTR models, samples of automatic transcription from multiple Latin manuscripts, a palaeographic "character map" and a list of manuscripts of the sort we are studying: https://github.com/parisbible. A versioned dataset of Paris Bibles found is available at Zenodo https://doi.org/10.5281/zenodo.7274507 (Wrisley and Guéville, 2022).

| In the manuscript Louvre Abu Dhabi, LAD 2013.051 | "Normalised" | "Semi-diplomatic" | "Diplomatic" |
|---|---|---|---|
| 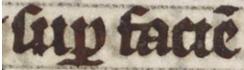 | super faciem | ſup*er* facie*m* | ſup faciē |
| 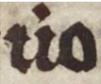 | vero | u*er*o | úo |
| 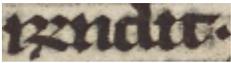 | respondit | *respo*ndit | R̃ndit |
| 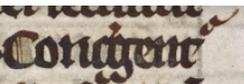 | congregentur | Cong*r*egen*tur* | Conǵgent˜ |
| 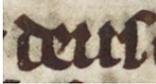 | Deus | ðeuſ | ðeuſ |
| 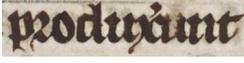 | produxerunt | pɹodux*er*unt | pɹoduxunt |

**Table 1:** A table illustrating sample words from a Parisian Bible at the Louvre Abu Dhabi, LAD 2013.051 and sample transcriptions: normalised, semi-diplomatic and diplomatic.

The second column (labelled "semi-diplomatic") shows how special letter-forms can be preserved as they are written in the text, making the difference between u/v or s/ſ, preserving the original capitalisation or spacing as much as possible. This method expands the abbreviations but usually indicates their purposeful expansion by the use of italics (or other methods such as underlining). It is a hybrid model from which we can understand editorial intervention in human reading, but is not amenable to plain text processing approaches which lack italics to encode such interventions (Widner, 2017). The last column (labelled "diplomatic") - and the least commonly used - seeks to preserve as much information as possible from the original manuscript. Similar to practices in epigraphic transcription, this kind of diplomatic transcription identifies written characters, linking them to Unicode "without spaces, punctuation or diacritics (unless these are in the source document), and without restoring lacunae or expanding abbreviations" (Bodard, 2021).

While there is a tradition of editing diplomatically, most editions of medieval texts have not been diplomatic, perhaps for pragmatic reasons. For dealing with the many letter-forms in medieval manuscripts that differ from modern alphabets (long s, insular d, rotund r, etc.), editorial traditions have differed about how to handle them as well. For the most part, there has been a rough convergence on a set of Unicode characters which stand in for "special characters" found in written documents. In the case of medieval manuscripts, transcribers employ abbreviations that have been the object of a considerable amount of study in palaeography, which initiatives such as the Medieval Unicode Font Initiative (MUFI) have rallied to describe (MUFI, 2015). Although most MUFI characters are in the public domain of UTF-8, meaning no other characters can be assigned to that specific UTF-8 code, one of the key weaknesses of the MUFI, in our opinion, is the nature of a so-called private domain of MUFI which includes less frequent abbreviations and letter-forms. Even though the character set can be downloaded and used with specific fonts, it requires workarounds for proper rendering and can be problematic in plain text workflows.[4]

Transcription norms have been designed as a way of ensuring the consistency of a critical text to be set down in print technologies which we associate with rigour and orthography. They ensure the accessibility of the critical text: for modern literacies and expectations of scholars and students alike. They give scholars access to a deeper understanding of the textual tradition, while ensuring ease of professional reading (Siemens *et al*., 2009). They both influence the kinds of research we might do and limit what we understand as possible. We would like to suggest that normalisation of textual features in the print-centred mentality of transcription entails a loss of information that could very well be useful--even central--to future approaches to digital research in medieval studies.

It goes without saying that the digital turn in the historical humanities has multiplied the different ways that we can access, create and use text. For some years, it has been possible to read and compare digitised medieval manuscripts at a distance and on a screen, a process which the International Image Interoperability Framework (IIIF) has sped up significantly. Access is not, however, only a question of the delivery of archival materials to far-flung parts of the world. The availability of digitised images of manuscript materials also allows us to access the text within them by creating text which goes by many names: machine-readable text, digital text, machine processable text, automatic transcription, or simply transcription (where automation is assumed). By accessing text, we mean more than just having a digital facsimile on the screen for human reading: we believe that access implies the possibility of creating a transcription which can be put to some specific, open, scholarly use.

The history of normalisation of transcriptions, we believe, has been closely linked to the evolving research landscape. While the first editions tended to preserve the micro features of manuscripts, the tendency to normalise transcriptions of medieval texts came about with the development of literary research. Whereas normalised editions allowed a community of researchers to have a common body of readings, scholars became interested not only on the "substantives" as defined by Greg (1950-51), that is the words themselves and their meanings, but also on the "accidentals", including the spelling, punctuation, or word-division. This change in research paradigm—going back to the text in the documentary record—impacted transcription schemes, in turn, facilitated by the evolving technologies available to the medievalist.

---

[4] This leads us to our project's tentative conclusion that for maximum flexibility across many platforms, the private domain of MUFI is to be avoided when dealing with HTR.

Journal of Data Mining and Digital Humanities  
ISSN 2416-5999, an open-access journal

http://jdmdh.episciences.org

## 2.2 Changing Technologies and Levels of Transcription

Let us return to the question of diplomatic transcription and the ways it has been used in the field as technologies for medieval studies have developed. Scholars of pre-modern cultures often speak of a diplomatic transcription or a diplomatic edition, in which characters are recorded as they appear with minimal editorial intervention or interpretation. Debates have focused on varieties of transcription, allowing for different amounts of scribal information to be captured. We see these debates about diplomatic transcription as connected to late twentieth-century critical practices of understanding documents within the context of their production and recopying, practices that emerged in the debates around New Philology and before, especially with the possibility of the delivery of digital versions of manuscripts in the form of images on the web (Rigg, 1983).

It might be argued that there are as many forms of diplomatic transcription practices as there are textual traditions. Critics assign different terminology to the "levels of transcription" (Robinson and Solopova, 1993), which sometimes overlap but also create subtle distinctions between transcription styles based on different editorial practices. Our purpose in raising this point is not to decide once and for all how diplomatic transcriptions should be done, but rather to suggest that *one's choices for encoding must arise not only from the specific textual scenarios at hand, but also from the ways that one wants to use such text downstream*. If the use for such text is a screen-based, documentary digital edition for scholarly reading, perhaps the transcription can be as specific to the textual tradition as desired and as specialised as the audience intended for the work. On the other hand, if the goal is to work with contemporary computational approaches to text, both the consistency and the concision of transcription norms become all the more important, since decisions made in both ground truth creation and subsequent model retraining embed bias within machine learning. If one aims to do both, a coordinated workflow must be designed to link the two together.

Let us consider some of the implicit assumptions of diplomatic transcription in some projects. Many, if not most, scholars make the distinction between semi-diplomatic and diplomatic transcription, the former usually expanding the abbreviations and the latter transcribing the text as it appears on the page, but others create more granular distinctions. In "Guidelines for Transcription of the Manuscripts of the Wife of Bath's Prologue" for the Canterbury Tales project (1993), Peter Robinson and Elizabeth Solopova defined four levels of transcription: **regularized** ("all manuscript spellings are regularized to a particular norm, perhaps the spelling of a manuscript considered authoritative"); **graphemic** ("every manuscript spelling is preserved (as: 'she', 'sche') without distinction of separate letter-forms as in a graphetic transcription"); **graphetic** ("every distinct letter-type is distinguished (as: r 'short' is transcribed apart from r 'round' and r 'long descender', etc."); and **graphic** ("every mark in the manuscript, every space, is represented in the transcription, even to the point of decomposition of letter-forms into discrete marks").

It is useful to note that Robinson and Solopova were limited by the technology they had access to, which in turn, impacted the kind of research questions they were able to address. Since they transcribed everything by hand, transcription of special letter-forms and abbreviations would have been too time-consuming to do across the entire manuscript tradition. As they mention, concerning the practicality of graphetic transcription: we found that while there seemed no cost in time in distinguishing these letter-forms in this first transcription, there was a marked cost in accuracy. It appeared that the concentration by transcribers on distinguishing these few characters meant that gross errors elsewhere in the transcription went undetected (Robinson and Solopova, 1993).

The situation with machine learning has completely changed our way of working, the time required for transcription, and the possibility of high accuracy. Since we do not transcribe everything from scratch, we can correct what HTR has already produced. Therefore, it is possible to focus on micro-features because most of the transcription is automatically done with a relatively low error rate. Given the kind of resources we have nowadays, namely artificial intelligence and HTR, we suspect that Robinson and Solopova would have probably chosen another transcription level, graphetic instead of graphemic, perhaps with characteristics from the graphic level.

In the Paris Bible Project, the transcriptions we have made of Latin Bibles could be characterised as a mixture of the graphic and the graphetic levels: we make the distinction between every letter form, we represent abbreviations, capitals, spaces, punctuation as faithful as possible, but we do not represent every single difference in the letter-forms (for example, longer or smaller vertical strokes, length, breadth or weight).

In the "Menota" (Medieval Nordic Text Archive) project for machine-readable editions of medieval Nordic texts, they distinguish three levels of transcription: normalised, diplomatic, and facsimile. They define them as follows:

- "A **facsimile** level (<me:facs>): A letter-by-letter transcription with a selection of palaeographic characteristics and the retention of abbreviations as in the manuscript.
- A **diplomatic** level (<me:dipl>): A letter-by-letter transcription with a small selection of palaeographic features and the expansion and identification of abbreviations.
- A **normalised** level (<me:norm>): A transcription in normalised orthography."

According to this classification, the transcriptions of Paris Bibles we have made would be described as "facsimile," although we do not encode all the properties they describe, e.g. unclear readings, erased and/or corrected text, initials, and *litterae notabiliores*, or headings.[5] In the case of the Menota project, the diplomatic level is described as an accurate transcription letter by letter in which the abbreviations are expanded and the number of palaeographic features reduced. As they explain, "a diplomatic transcription, [...] requires more editorial intervention than the facsimile transcription in the form of the interpretation of abbreviations and the normalisation of allographic variation." The resulting transcription will be more readable than the facsimile transcription, the latter being perhaps more useful for computational analysis than for scholarly reading.

---

[5] Inline mark-up could, of course, be used as a complementary process for making these observations in the HTR system we used, Transkribus.

The two examples described above (the Canterbury Tales Project and the Menota project) are representative of different ways digital projects in medieval studies have approached complex transcription. At one end of the spectrum, the Canterbury Tales project uses transcriptions which may be called strictly diplomatic, in which every feature which may be reasonably reproduced in print is retained, not only spelling and punctuation, but also capitalisation, word division, and variant letter-forms. The layout of the page is also retained. Any abbreviations in the text will not be expanded, and, in the strictest diplomatic transcriptions, apparent slips of the pen will remain uncorrected. Such editions are often so close to the originals as to be too complex for the nonspecialist reader, or in any case no easier to read than the originals. At the opposite end of the spectrum in the Menota project, there are fully modernised transcriptions, where the substantives (Greg, 1950-51) are retained but everything else is brought up to date, in some cases to such an extent as to make it questionable whether they are to be regarded as transcriptions at all. In between these two extremes, a number of levels may be distinguished— 'semi-diplomatic', 'semi-normalised', etc. —depending on how the accidents of the original are dealt with.

While it is true that development and use of TEI-XML for the purpose of scholarly editions most definitely had an impact on the way scholars transcribe multiple layers of texts (Driscoll, 2006), we believe that a near future of medieval studies will include significantly more automatic transcription than is found at present, requiring scholars to find dynamic ways of capturing and organising multiple levels of data. It is not so much that automation will replace human labour, but the former will make the latter more scalable and customizable and will open up intermediary forms of textual exploration between browsing the digitised manuscript library and preparing a full edition. In our opinion, when the question of automation is combined with transcription, it is only logical that scholarly infrastructure for dealing with different layers of transcription will be created, opening new pathways for text creation and analysis. Medievalists will no doubt be working with these different kinds of transcriptions, so modes of encoding such as TEI-XML seem particularly appropriate for collating the different representations of the same document, with different uses that could be made of them by machines and humans. For some research, such as in the Paris Bible Project, only a diplomatic transcription is needed for computational analysis, but this does not preclude the creation of a parallel normalised version for human reading or for cross-version querying using TEI-XML or other means. Indeed, much more thought needs to be given to the ways in which such levels of data created about documents in the iterative HTR process can be incorporated into the archive for use and reuse by future researchers.[6]

One of the most common uses for HTR-generated text at present seems to be searchability of archival documents (Stutzmann *et al.*, 2018), although the critical literature describing the use of such technology is expanding quickly (Nockels *et al.*, 2022). There are, of course, other possibilities when we move closer to traditions of complex analysis and interpretation of texts well known in medieval studies. For example, one might want to have an automatic transcription as a draft baseline for creating a new documentary edition, or creating a diplomatic layer for a new critical edition. Likewise, an unedited, but searchable, transcribed text could be used for semantic annotation or for the purposes of genetic criticism. In a complex textual tradition, any number of transcribed witnesses might be made for comparison, alignment or higher-level analysis (Jänicke and Wrisley, 2017). Each of these end goals is facilitated by transcription, but the nature and norms of that transcription impact the extent to which we can accomplish our "scholarly primitives" with ease. Diplomatic transcription adds new kinds of information to corpora specific to the scholarly questions underlying them. In an age of rapid advancements in machine learning for computer vision it may be time to re-theorize the diplomatic transcription, or replace the term altogether. It is quite possible that in a few years' time the global community of medievalists will end up with a mass of automatically transcribed text that is quite difficult to compare computationally.

---

[6] One such conceptualization of rich data linked with texts has been elaborated by the Cadmus framework. See Fusi (2018).

In the creation of ground truth data for training HTR models, there are many choices to be made and advice is generally quite vague: *transcribe as closely as possible what we see in the document*. Our own approach to diplomatic transcription is linked to purpose; we chose a transcription scheme in line with the kinds of information we want to represent, and this information is available to us for analysis in the resulting text. It is important, however, to qualify the expression "what we see," because sometimes a transcriber is confronted with a passage in which our knowledge of the ways that the platform tends to react changes the way that we transcribe. The socio-technical elements of ground truth creation have been described by others, so we will not rehearse them here (Alpert-Abrams, 2016; Cordell and Smith, 2018). How can we mitigate these problems from the beginning in the design of a HTR-enabled text creation project, especially if our ground truth and HTR models are made publicly available and will likely be used and adapted by others? (Romein *et al.*, 2022). How can we be sure that complex modes of transcription–encodings in their own right–are machine processable and that they do not interfere with basic downstream processes, such as tokenization and word counting?

**III. DESIGNING CORPORA FOR TRANSCRIPTION**

Transcribing for modern readers so they can read the text without difficulty and transcribing for a machine are two very different tasks. Computational linguists have been calling for the "representation of manuscript reality" in medieval corpora for some time (Honkapohja *et al.*, 2009) by encoding linguistic, palaeographic, and codicological features in digital editions. This approach, of course, sees value in editing pre-modern texts, but wishes for them to be available in "unadulterated form" so that their non-normalised complexity can be also used for research purposes. Creating an automated transcription of a manuscript is not the same as digitising it, rather it is creating an imperfect representation of it–with all the limitations of any computational model–in order to be able to examine that text through specific lenses. Transcribing for a machine, however, does not preclude the eventual editing of works, but let us speak first about creating actionable transcriptions for computational reading.

The abbreviation in medieval studies is typically framed as a skill of medieval literacy (and a medievalist's literacy) in order to be able to read in manuscript. Although there have been some quantitative studies about abbreviations in Latin and vernacular languages (Bozzolo *et al.*,1990; Hasenohr, 2002; Römer, 1997), they are usually something to learn, understand, decode and then encode or uncollapse, usually when making a transcription or an edition (Honkapohja, 2013 and 2018). Research tools exist for understanding them, such as Cappelli's *Lexicon Abbreviaturarum*, the famous white reference book on the medievalist's shelf (and even available in digitised forms, Ad fontes; Abbreviationes Online). In thinking through our research process of transcribing medieval manuscripts using Transkribus, it occurred to us that most researchers do not use abbreviations as features that can be useful in and of themselves but rather as a "problem" to resolve in order to understand the meaning of the text, even to mark as an expansion in a critical edition indicating how the editor has interpreted the abbreviation.

We believe that theoretical potential of working with diplomatic transcriptions is different; it is not a stand-in for close reading in manuscript, rather it facilitates textual analysis with the help of computational tools. In our Paris Bible Project, we are interested in countable micro-features in copies of the many Paris Bibles extant in the world in order to perform predictive analysis about the copying and scribal habits, dare we say styles, of such Bibles (Herrmann *et al.*, 2015). For designing a transliteration scheme, turning to Cappelli for examples of the most common abbreviations and letter-forms in particular manuscripts was not particularly helpful. The latter is an all-purpose reference work aspiring to spatiotemporal and generic breadth. In our case, we are working with one domain and a relatively constrained space and time in which it emerges. We encountered similar issues working with MUFI, including the fact that most of its character set does not appear in our manuscript. In sum, if we consider any transcription system as a form of encoding medieval scribal data, our approach is to choose one pragmatically which corresponds to the realia of our specific corpus of manuscripts.

To train an HTR model for the hands of Paris Bibles, we first needed to create a transcription of a handful of folios. To do so, we identified about 40 special characters used as abbreviations: superscript characters which are placed on top of letters, the so-called "combining letters" of Unicode (¯; ´ ;¨; ' ; etc.), some special characters (p; 2; p; 7; q etc.) and special letter-forms (ſ; ð; ꝛ) to distinguish from their common form (s, d, and r); and finally; superscript letters ( ⁱ ; ᶜ ;ᵐ; ˢ ; etc.). The first and the third groups can be in used with many letters and indeed we discover new combinations on every page we transcribe. The quantity of unique abbreviations and letter-forms is, however, somewhat limited. We opted for a pragmatic, adaptable Unicode solution that works easily with the HTR system. Of course, there were slight palaeographic variations between specific letter-forms and in the placement of punctuation, a fact that became particularly apparent as we moved across different manuscripts and different contexts of manuscript production. This fact did not lead to the multiplication of new forms of encoding, but rather sets of problems which we resolved in the project guidelines.[7]

---

[7] Since the guidelines for transcription in our corpus are evolving as our thinking about the corpus and the different manifestations of the Paris Bible in Latin is refined, we have chosen to publish (and version) our norms at the project site, including the special characters used in the project: https://parisbible.github.io/guidelines/. The list of special characters is accompanied by snapshots of sample letters in manuscript.

# IV. SOME BASIC PRINCIPLES OF TRANSCRIPTION OF MANUSCRIPTS FOR HTR

Each project, each manuscript, each hand being different, the list of possible Unicode characters to be used for transcription can evolve as the knowledge of the corpus does. Accordingly, we suggest that there is not one definitive list of guidelines that can be used and applied to any single project, and especially not to all projects, but rather from a large number of possibilities, case-specific criteria must be designed in line with project objectives. Like the case of editing texts itself, inevitable questions of data loss arise when we imagine transposing the writing found in manuscript to letters in a digital writing system. Establishing a specific list of special characters, and setting principles to follow is fundamental, but every attempt raises multiple questions: How much information do we include? How do we encode variance, exceptions, and aesthetic scribal habits? How do we choose what is an accepted variant to be encoded, or alternatively, how do we decide when a variant is not significant enough a feature to merit attention? How do we know what forms of encoding are enough to do our documents justice? How do we prioritise the principles in the case of contradiction?

In what follows, we outline some basic principles for transcribing for computational reading which have emerged at this stage of our research. The list of principles below is not meant as the last word on the subject, but as a touchstone for other projects with similar issues. We invite its revision through scholarly debate as the community works with archives across languages and periods.

## 4.1 Principle 1

Although transcribing for machine learning is fundamentally an interpretative activity, the **first principle** to abide by should be that the transcription must be as close as possible as what you see in the manuscript. Closeness is admittedly a vague notion, since any choice of Unicode characters would not suffice to render all the palaeographic variety from scribal hand to hand. We might restate the first principle as such: if there is a basic character in Unicode which corresponds to what you see, and that letter exhibits insignificant variance across your document for the research problem at hand, there is no reason to opt for a more complex character encoding. This principle, of course, requires us to think before we transcribe, or even revise our initial transcription systems iteratively, rather than relying on pre-set characters recommended by a print editor or by editorial guidelines made for paper editions. Even with such a basic principle of using a corresponding Unicode character for the letters you see in manuscript, there is the potential from project to project, or manuscript to manuscript for contradiction.

In some projects, it might also be useful to use a digital tool such as ChocoMufin (Chagué et al., 2021, Pinche and Clérice, 2021) to mitigate against inconsistencies in the transcription. While ChocoMufin provides a useful check, it also tends to normalise the transcriptions, a fact which the creators acknowledge (Chagué et al., 2021). It is obviously better to establish guidelines before transcribing, if possible, rather than waiting for the moment of ingestion into any repository, but with such a complex process carried out by many researchers, it is inevitable that guidelines will vary slightly. In the case of the Paris Bible Project, we did not use such a tool to check the quality of the training set, but fully hand-corrected the ground truth against a set of authorised characters. While not scalable across a corpus of many different genres, such an approach matched the relative consistency of Latin Bible scripta in the thirteenth and fourteenth centuries. Whatever guidelines are ultimately adopted by a project the character set used to train models must be made explicit.

### 4.2   Principle 2

When transcribing any document, since there is always the possibility of variance in portions of the document you have not yet seen, the **second principle** is that it is useful to have a preliminary "scan," even a random check of different parts of the full document you want to transcribe automatically, or through samples of the corpus you will be working with, before beginning transcription. A first pass of transcription in the early stages of a project, or when moving from one manuscript to another, allows you to create a working list of special characters. It can be useful to remember that the list of special characters which can be used for any given system is finite, so it will be important to prioritise the ones which are the most significant. Finalising the special character list may just be iterative. It may be important to revise the list of special characters over time when moving to samples of documents which are quite different, or say, across genres in the same language or documents written by very different scribes. One way of assessing whether palaeographic variance is significant enough for two letters to be encoded with separate characters once initial automatic transcription has been done, we have found, is to examine the spelling in the manuscript using full text search.

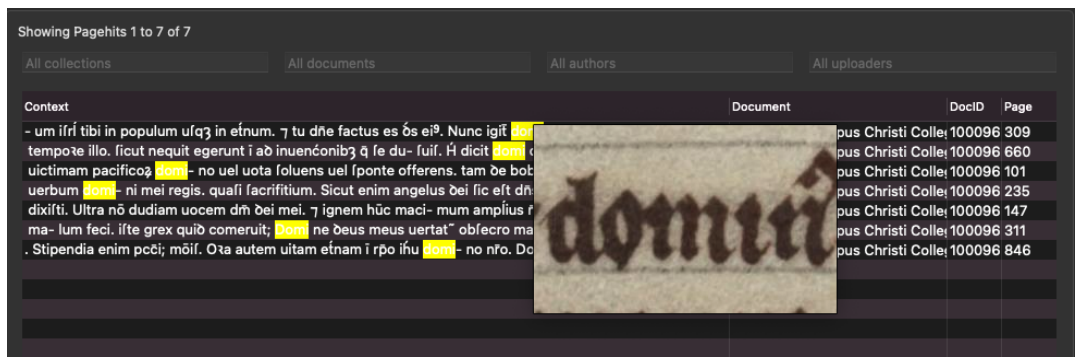

**Figure 2**: A screenshot of a full text search of the word "dominus" in a partially transcribed document, Cambridge, Corpus Christi College, Manuscript 049. Visualised in Transkribus.

### 4.3   Principle 3

Since there is inevitable variety in the hands or even in a single hand, a **third principle** related to the second is to take care when encoding with fine granularity the "aesthetic" quality to some graphemes (in our case, this meant questions like spacing, or the letters v and p), or a variance in the placement of certain abbreviations (for example, the macron) which create too great a variety of encoding choices or difficulty in ordering the characters in the transcription. This principle will seem to some as foreclosing on some of the palaeographic singularity of manuscripts, but from the perspective of computational analysis based on frequency, a very large variety of singular occurrences of a letter would bear less meaning.

### 4.4 Principle 4

Since in our research we are creating machine-readable transcriptions for the purpose of computational analysis, the **fourth principle** we posit is not to employ proprietary fonts (such as the proprietary subset of MUFI characters) that pose character encoding problems or are complex to display across multiple tools in a pipeline. For most of the private domain of MUFI other UTF-8 codes can be assigned to other characters. Whereas this principle might seem to contradict the first of transcribing what you see, the idea is to choose characters which are sufficiently like what you see within a "minimal" set of choices of characters.

### 4.5 Principle 5

The **fifth principle** is not to make transcription guidelines that will add features to the transcription that can be easily "undone" by common forms of digital analysis. Put another way, basic natural language processing (NLP) pre-processing tasks such as lowercasing, tokenization or removing punctuation need to be considered when training HTR to transcribe text. For example, imagine that there are different versions of the letter "d" in a manuscript; it would not be a good choice to encode one of those variants of d with a capital letter "D" since the routine lowercasing of text with the Natural Language Toolkit (NLTK) would erase a carefully encoded distinction. Likewise, tokenization–the process of dividing text into strings–might divide words which are encoded with an apostrophe, a blank space, hyphen or brackets, so care must be taken to avoid using such common delimiters to encode particularities of language from manuscript. This is also important when thinking about so-called "smart" HTR models which attempt to train HTR to produce beyond what is seen in manuscript, for example to transcribe an abbreviated word with two layers of data, both the abbreviation and the expanded form of the word (Rabus and Tikhonov, 2022).[8]

### 4.6 Principle 6

The **sixth principle** implies that transcription choices depend on a deep knowledge of the source and its production. Spaces and punctuation are highly variant, as are the text blocks in manuscript. In our specific research, the importance of marginalia, predominantly corrections, running titles and catchwords, is negligible for our research question and so we do not transcribe them in ground truth or include them in page layout. We do not mean to imply that marginalia are not useful as data, but in the case of the Paris Bible Project, we focus on the main text blocks; examples of corrections or marginalia were often in a different hand. Other objects, such as glossed bibles, have a more complicated layout with many text blocks intertwined and other medievalists might want to consider this additional data. Each project being different, scholars need to ask themselves: what is the purpose of the project and, what do we need to focus on, or what is relevant? Once these relevant matters have been identified, the most significant ones can be included in the principles of transcription.

---

[8] The general principle of using different forms mark-up as complementary to character encoding for helping to distinguish word forms, and for model training, is one that we suspect will see much more reflection in years to come.

### 4.7 Prioritising the Principles

In the case of contradictory decisions, we add the following coda to the principles outlined above: *there is a need to prioritise the principles based on the research questions at hand*.

For knowing where to place the abbreviations (such as the macron) in the sequence of letters of transcription, we made an arbitrary decision to consider what letter the abbreviation replaces, rather than where it is placed. Let's take the example of the word "bien." In the manuscript BnF français 24428, the scribe wrote it in ways that can be transcribed either biē or bīe, the macron being often written on top of both letters, sitting in the middle of them. That is to say that transcription is never fully divorced from a scholarly interpretation of what was created by the scribe.

In the case of spaces, punctuation and special letter-forms, a larger sample should be consulted. What we mean by this is that norms of spacing, the formation of letters, even dividing letters on a manuscript folio do not come close to the exigencies of normalised print culture. It is worth looking at the overall practice of a scribe or scribes when making decisions about transcription. For the purposes of creating ground truth for HTR transcription, a single occurrence of a script phenomenon is best thought as an exception rather than a principle, until it can be understood in the larger trends of what a particular scribe does. How we decide on that larger sample depends on the scope of any particular project. It is not recommended to create a new Unicode or a spacing decision based on a single example, but rather on a larger sample.

These principles outlined above are not universal, rather we have found they are enriched by work across different domains, periods of time and languages. Again, we have focused on principles which allow us to create philologically faithful transcriptions for textual analysis where our research question aims to understand scribal input. It follows from these principles that a general model for transcribing the entirety of medieval Latin, or French or Arabic is unlikely for philologists interested in using HTR, but instead a variety of sub-models is a desirable goal for scholars belonging to a language-specific community. For larger more general use, such as keyword searchability in libraries or archives, such levels of precision with transcription systems may prove to be unrealistic for now, and a slightly more normalising approach may work better.

# V. BUILDING A DATASET OF MANUSCRIPTS FOR TRANSCRIPTION

## 5.1 Choosing a Corpus of HTR Transcriptions

In this section we would like to offer some practical suggestions for thinking about building a dataset of transcriptions from one or more manuscripts. One reason for working with manuscript-level transcriptions is that there is something about the scribal behaviour in the manuscript(s) which is worthy of critical attention and that can be detected from the text. Purely palaeographic variation would not be served well by HTR, as the goal of automated transcription is to convert a document into machine readable text, and so only very different letter-forms would be thus captured. Another key point when considering automatic transcription of manuscripts is that our ability to build a custom HTR model is usually based on the manual transcription of a few dozen pages of text (several thousand words). This step of the process can be a very time-consuming one in medieval languages and any project should budget enough resources to get started. If existing transcriptions or an edition exist, it is important to remember that documentary style editions are better than composite critical editions; moreover, to adhere to the philologically faithful approach outlined above, they will have to be "un-normalized" to be used as ground truth for model training.

The performance of models depends on the neural network as well as the quality and quantity of ground truth, and Chagué *et al*. (2021) plead for a pooling of resources, or in this case, of ground truth. They argue that there are two ways of training models: from scratch or by refining a pre-existing model (as we did with Hodel's). However, most projects are isolated: they create their own ground truth and rarely share it openly. Although generic models for niche historical languages seem very unlikely to be efficient in the foreseeable future, sharing a project's ground truth with the transcription statement will support the creation of new models based on similar research questions, scripts and/or transcription rules. Having access to open ground truth has been a struggle that initiatives like HTR-United try to address by allowing anyone to share HTR datasets with standardised metadata. We share descriptions of our ground truth on HTR-United and the ground truth used to train models, plus selected transcriptions on GitHub.

One can most certainly work with transcription in a single manuscript, but in this case, the majority of the codex, or the part of the codex of interest, should be in the same hand and should be long enough to justify the start-up time of preparing the training data. If the end goal is not to have a long portion of the manuscript (or multiple manuscripts) transcribed, it is probably best simply to transcribe by hand. It is important to remember that when applying one HTR model trained on one hand to a different hand, the retraining process can be significant. The same can be said for adapting a model from one scholar's project to another. An example might illustrate this last point best. In Table 2, at far left we illustrate a normalised Vulgate passage from the beginning of 3 Kings (the medieval designation for 1 Kings). Using a public model in Transkribus, we can transcribe the second column with minimal effort after layout analysis. In the third and fourth columns are visualised transcriptions after correcting a set of sample pages and retraining the model. With time and effort, there is a gradual, although not perfect, convergence from one project's model with specific purposes to another's.

| Normalised Vulgate | Gothic_Book_Scripts_ XIII-XV_M4 (Hodel) | LAD 1.0 (Paris Bible Project) | LAD 1.3 (Paris Bible Project) |
|---|---|---|---|
| Et rex David senuerat, habebatque aetatis plurimos dies: cumque operiretur vestibus, non calefiebat.<br>² Dixerunt ergo ei servi sui: Quaeramus domino nostro regi adolescentulam virginem, et stet coram rege, et foveat eum, dormiatque in sinu suo, et calefaciat dominum nostrum regem.<br>³ Quaesierunt igitur adolescentulam speciosam in omnibus finibus Israel | e rex dauid senuerat habebatg letatis plit umos dies Cumaz opt rireturu stibʒ no ca letiebat •<br>D veerunt ergo ei sern sii •<br>Queramus dno noo rege adolescentulam iurgine • e stet coram rege • et foueat eu • dor miatqʒ in sinu suo et calefa ciat dum nrm regem •<br>Quest erunt ergo adolescentulam speciosam • i omib fuubʒ isrl | Croc ðauið scnucrat habcbatq ctatis plu- imos ðics Cumqʒ opÍ ructur u- stibʒ nō ca- sciēbāt.<br>Dfġġtunt cigo q sip sui.<br>Qucramus ðño npo rigi aðolescentulam uuginē; 7 stcī coʒam rege; ct soucat eū. ðoʒ miatqʒ in sinū suo ct calcta cat ðn̄m npm rogem.<br>Ouesi crunt crgo aðolqscntulam spocosam. ī onm̄ibʒ siaibʒ isiÍ | Cux ðauið sciucrat habebatqʒ letatis plu- unos ðies Cumqʒ opi rireturɡ u- stibʒ nō ca- leliebat.<br>Oixesunt ergo ci seciu- sui.<br>Queramus ðño nr̄o regi- aðolescentulam uirginē; 7 stet coʒam rege; et foueat eū. ðoʒ miatqʒ in sum suo et calefa- ciat ðn̄m ntm̄ regem.<br>Quesi erunt ergo aðolescentulam speciosam. ī om̄ibʒ finibʒ isrÍ |

**Table 2:** Sample transcriptions of the beginning of 3 Kings (modern, 1 Kings) from Free Library of Philadelphia, Rare Book Department, Lewis E M 063:01-31, folio 63r, using three different models.

Since medievalists working on an edition are often working with a number of different witnesses which exist in different countries and which were copied during different periods, the question can arise whether HTR is an appropriate method for their transcription. A team needs to ask itself if the different witnesses are found in different hands, does the time allotted permit training a model from scratch or adapting a separate model to each one of the various hands? Additional questions about access to resources are also worth asking. Is there access to a downloadable scan of the document or a IIIF manifest from a digital manuscript library? Is the document scanned in its entirety or is the scan partial? Is the quality of the digitization satisfactory for your purposes? What are the laws around the reuse of images across institutions and countries? Finally, since the images will sit on external servers, does the holding library allow you to upload their images for such research purposes? If you share ALTO XML for images which are not in the public domain, how will such data be reusable?

Defining the scope of the project as well as the type and number of manuscripts to be used constitutes the first step of the process, but it is one which needs to be carried out with a realistic understanding of what HTR is able to achieve. In the Paris Bible Project, gathering digitised manuscripts, assembling and labelling data was our first struggle, largely because of the way manuscripts are preserved and made discoverable. The history of collections and the way institutions describe objects, their approach to digitization, and their policies towards accessibility and reusability were all significant hurdles.

## 5.2   Challenges of Building a Corpus of Paris Bible Transcriptions

Not all medieval manuscripts, especially Paris Bibles, are digitised, or even discoverable. Some Paris Bibles occupy a symbolic role in historical manuscript libraries. They were originally objects intended for individual use–for studying, teaching or preaching purposes–but they have not traditionally made up a single medieval collection. To our knowledge, there have not been collections of fully digitised ones either. Today many cultural institutions with a medieval collection–be it a museum, a seminary or a library–possess at least one Paris Bible. There are significant numbers in private collection. Since they represent an important moment in the history of book production and in the history of devotion, preaching and teaching (Light, 1994; Light, 2012; Ruzzier, 2010), they have become a "must-have" for many collections, prestige objects that are, in some cases, among the very first numbered manuscripts. In other collections, on the other hand, they can be notably undiscoverable, considered as a "common" object, even difficult to find. This paradoxical situation of Paris Bibles makes them both numerous, but complex to work with. It is also difficult to work with them since their digitisations are of unequal image quality, they are not all publicly available and they have competing or contradictory library catalogue descriptions. The term we have used (Paris Bible) is not universally employed, and such documents are found instead under their specific terms in various European languages: *Biblia sacra, Pariser Normbibel, Bibbia dell' università, Universitetsbibel*, or even under the very general denomination "Bible." Given this diversity of nomenclature, we have relied on manual and visual identification, using discipline-specific rich metadata about these objects, primarily in European and North American collections that adopted early integral digitization.

The challenges described above in the constitution of one's corpus within the constraints of what HTR does well are important to consider when designing a research project involving transcription of medieval manuscripts. What makes the Paris Bible a strong candidate for HTR transcriptions is the fact that they are usually written in a somewhat standard Gothic hand. A project which wanted to use HTR to transcribe the manuscripts of a large textual tradition such as the *Roman de la rose* or the *Speculum historiale* would not enjoy the same success, due to the inevitability of difference in hand across the various witnesses.

### 5.3 Problems of Metadata

Metadata are important when it comes to transcriptions made from manuscripts. They are what allow us to group together different artefacts. Working with medieval manuscripts is difficult, however, as we mentioned before, since cultural institutions describe their objects in vastly different ways. The material evidence which can be gleaned from well described manuscripts can be enormously helpful for contextualising the transcriptions we are making. A specialised library with a significant manuscripts collection has more chances to have a specialised curator or cataloguer who would provide many codicological and artistic details, including, for example, the justification size, the number of lines, the ruling or presence of catchwords. On the other hand, a museum with a variety of objects and a normalised description process used for ceramics, paintings or manuscripts would be much less specific. Moreover, countries and institutions use different norms for writing dates or locations as well as different languages. Overall, we had to treat the metadata with caution.

The quality of legacy metadata is fundamental to our project: our process is recursive, which means that we carry out transcription, correction, retraining and analysis of different datasets. We needed to start with well-known documented manuscripts which could serve as reference points for understanding the textual tradition. In our specific case, the more that we know about the time and place of the copying of a manuscript and the different hands found in it, the more we are able to "predict" about others. There was an inevitable degree of normalisation across different libraries and national traditions which allowed for us to compare like objects. It is very possible that large research projects working with many manuscripts encounter a similar phenomenon. Lastly, but connected to the question of metadata, we needed to make choices regarding the manuscripts used for training purposes. Our model was first trained on one single manuscript, the Bible (LAD 2013.051) from the Louvre Abu Dhabi collection (Guéville, 2021), and not even in its entirety: we used the text of Genesis, part of the Exodus and the books of Matthew and Mark. Prologues, marginal corrections, and other Biblical books have been completely ignored in the first phases of the project, thus creating a potential bias toward specific words contained in these books. We then added a handful of manuscripts using the same texts. Our latest "composite" model is currently based on a dozen manuscripts, which are not representative of all locations, dates and traditions.

18Journal of Data Mining and Digital Humanities  
ISSN 2416-5999, an open-access journalhttp://jdmdh.episciences.org

## VI. DESIGNING HTR MODELS FOR RESEARCH QUESTIONS

### 6.1 Training a Model for Transcribing Medieval Manuscripts

The steps to actualising an HTR model for transcribing medieval manuscripts extend beyond understanding what kind of transcription scheme is most appropriate and what digitised manuscripts are available. The process of training a HTR model involves a non-trivial layout analysis step in which baselines and polygons are identified and with which a given ground truth transcription can be aligned. Even when a manuscript is available via IIIF, sometimes the quality of the image does not allow for automated layout analysis. This situation arose with a well described and well localised Bible produced in Bologna (BnF NAL 3189) which we were particularly excited about using as a reference manuscript due to its origin and its known copyist.

In some cases, given the research questions of a particular project, full transcriptions are desirable, but not truly required. That is to say that a successful project with HTR transcription of medieval manuscripts depends on a large number of factors and one is almost inevitably required to make a compromise between the availability of specific texts, the specificity of scribal hands, the quality of digitization of the manuscripts and the tasks one would like to carry out with the resulting transcriptions. In modern or contemporary archival transcription projects, often an emphasis is placed on a model being able to transcribe a variety of hands with precision. With medieval transcription projects, the added dimensions of medieval textuality (multi-scribal composition, compilation, rebinding, marginalia, etc.) have to be taken into account for project design. In other words, the more specific the research question, or the more complex the sources, the more precise your transcriptions should be, and the more problematic the idea of a general model becomes.

In the Paris Bible Project, in order to create an HTR model, we had completed a number of steps: identification of our corpus of folios, normalisation of our metadata, and design of our transcription scheme. We trained our first model in Transkribus, LAD 1.0, based on the public HTR+ model "Gothic_Book_Scripts_XIII-XV_M4" (Hodel) to which we added data from the manuscript kept in the Louvre Abu Dhabi collection (LAD 2013.051). We created eight pages of ground truth, amounting to 588 lines and 3547 words. It cannot be understated that to arrive at the first model of a project, especially when this requires careful transcription from manuscript, is an arduous process. The human effort in the process grows lighter after these initial steps, with emphasis shifting from transcription from scratch to post-correction of HTR outputs and assembly of new manuscript samples. We trained two subsequent models, LAD 1.1 and LAD 1.2, with 16 more pages of ground truth (1592 lines and 9632 words in total) from the same manuscript. The characteristics and performance of the various models are summarised in Table 3. In sum, HTR model design is a time-consuming endeavour which connects specific research questions to a general understanding of the performance and characteristics of the HTR system, but also for which the results are somewhat difficult to predict in advance.[9]

Our working guidelines for transcription, a list of special characters used in transcription and the project blog are available on the project site: https://parisbible.github.io/. Our project GitHub includes ground truth samples for HTR model construction, samples of automatic transcription, a paleographic "character map" and a list of manuscripts used: https://github.com/parisbible.

| Model | Date | Base model | Train_set | Validation_set | lines | words | CER (training) | CER (validation) | Comments |
|---|---|---|---|---|---|---|---|---|---|
| LAD 1.0 | 15/08/2020 | Gothic Books (Hodel) | 7 | 1 | 588 | 3547 | 0.27% | 4.52% | Based only on the manuscript LAD 2013.051. Bias towards Northern France, second half of mid-13th century. Bias towards Genesis, Exodus, Mark and Matthew. |
| LAD 1.1 | 22/08/2020 | LAD bible 1.0 | 19 | 5 | 1592 | 9632 | 11.89% | 7.20% | Same bias as LAD 1.2. Based only on the manuscript LAD 2013.051. Same GT as 1.2 with a different base model. |
| LAD 1.2 | 22/08/2020 | Charter Scripts (Hodel) | 19 | 5 | 1592 | 9632 | 0.62% | 4.14% | Same bias as LAD 1.1. Based only on the manuscript LAD 2013.051. Same GT as 1.1 with a different base model. |
| LAD 1.3 | 26/10/2020 | Gothic Books (Hodel) | 30 | 9 | 2516 | 15258 | 0.51% | 3.01% | Based only on the manuscript LAD 2013.051. Bias towards Northern France, second half of mid-13th century. |
| PBP 1.0 | 29/06/2021 | LAD bible 1.3 | 16 | 9 | 152 | 8840 | 2.04% | 12.76% | Composite model based on Paris Bibles from around Europe in the 13th and 14th centuries. Bias of books, localisation and dates mitigated. |

**Table 3**: Statistics on the HTR models trained in the Paris Bible Project.

## 6.2 Towards a Composite Model?

Sometimes it proves valuable to combine different sources with different hands into a single model in the hope that the general result will prove more successful across a variety of documents. This approach can be useful for limiting the bias of a particular set of manuscripts when it comes to abbreviations or spelling and to avoid the problem known as overfitting, where assumptions about a known dataset are erroneously imposed onto a new unknown dataset. Since our first models LAD 1.0, LAD 1.1, LAD 1.2 and LAD 1.3 were developed using a very limited corpus[10], we soon realised their limits: they were heavily biased and often we found examples of their common abbreviations reproduced in transcriptions of new manuscripts where they were not present. In order to reflect the diversity of Paris Bibles and limit some of the bias introduced, we decided to increase the size of the dataset and train a new model based on multiple manuscripts, reflecting the multiplicity of traditions, locations and dates of production. We were aiming for a model with enough data to extrapolate from one Biblical manuscript to another.

---

[9] In the final stages of this article in early November 2022, it was announced that the proprietary and outdated code of the HTR+ engine with which we created the models mentioned in Table 2, is being phased out. We will be retraining models for Latin Bibles with abbreviations using the PyLaia engine.

We constructed a dataset of approximately 450 folios from 24 manuscripts[11]. These manuscripts were produced mainly in France and England, with examples from Italy, Germany and Spain (see Table 4) and the folios were selected from a variety of books of the Bible: 74 books out of 81.

| Origin | Number of mss |
| --- | --- |
| England | 7 |
| France | 6 |
| England or France | 1 |
| Germany | 2 |
| Italy | 3 + 1? |
| Spain | 2 |
| Spain or Italy | 1 |
| Unknown | 1 |
| Total mss | 24 |

**Table 4:** A list of the number of manuscripts used to construct our "composite" Paris Bible dataset with digitised copies from around Europe.

This composite dataset is a kind of "artificial" Paris Bible. We collected samples of folios from full and partially digitised Paris Bibles with a maximum of variables in mind (book of the Bible, dating and localization of the codex, codex size, etc.). We then trained a new HTR model (PBP 1.0) based on 3 corrected folios from each Bible manuscript. Had the Paris Bibles of the world been mostly IIIF compliant, the gesture would be even much easier, as with the HORAE project carried out on Books of Hours (Stutzmann and Chevalier, 2021). Instead, for us aiming to capture a diversity of Latin Bible codices has meant turning not only to copies which are already available in European and American libraries, but also to smaller, under resourced or isolated libraries, and even to legacy databases (such as Mandragore and Digital Scriptorium) which never intended to publish fully digitised codices. As one might expect, the results of this composite model were less than optimal when used to transcribe new unknown texts. What we learned in the process of creating this model is that while there are a number of well-dated and localised Paris Bibles in the world, the hands represented by those codices are quite diverse and the three folios we transcribed from each most likely did not provide a large enough sample to capture these hands. In addition, the quality of digitisation and the ability to carry out layout analysis was a significant hindrance to including the data from southern European libraries.

---

[10] LAD 2013.051, to which were added three manuscripts from the BnF collection, Latin 40, Latin 10421 and Smith-Lesouëf 19.

[11] The composite dataset were made up of excerpts from Berkeley UCB 12; British Library Additional 50003; British Library Royal 1 D I; BnF Latin 40; BnF Latin 10421; BnF Latin 10428; BnF Latin 14232; BnF latin 14238; BnF SL 19; Trinity Hall Library, ms.22; Cleveland Art Museum 2008.2; New York, Columbia University, Burke Library at Union Theological Seminary, UTS MS 072; Harvard Typ 446; Louvre Abu Dhabi, LAD 2013.051; Free Library of Philadelphia Lewis E31; Free Library of Philadelphia Lewis E32; Free Library of Philadelphia Lewis E37; Free Library of Philadelphia, Lewis EM 063; Library Company of Philadelphia 9; Madrid 12802; Arc Priv 3 Montecassino 2018; Schaffhausen, Stadtbibliothek, Ministerialbibliothek, Min. 6; Swarthmore College BS 75 (partial); University of Pennsylvania Ms 1065.

We can extrapolate from this experience that each time a new HTR model is developed with a new question in mind, a considerable amount of new data needs to be created, but also that the effort to create a diverse dataset is significant. The step between our initial model based on one manuscript and the composite one was quite large indeed. In summary, we are confronted with a paradoxical situation. On the one hand, we do not see a general model being a strong possibility in the near future, for generating high quality transcriptions, even for a manuscript corpus as "uniform" as that of the Paris Bible. This situation suggests that a number of models for subsets of the corpus is perhaps a better strategy. On the other hand, the number of digitised manuscripts from all the collections in which we find Paris Bibles is uneven, seemingly precluding the possibility of an extensive number of models.

**6.3 How Much is Enough?**

The field of machine learning is one which can be alienating to those with traditional humanistic training, since it introduces notions such as "ground truth," "training data," or even "gold standard" data which have traditionally belonged to the sciences. One other key idea in machine learning is that of "predictive analysis," again not one that is usually integrated into a study of medieval manuscripts. Predictive analysis can be tricky because we know that inaccuracies or biases in have the potential of creating false conclusions. Two questions which loom in the background of our work, and will be important for a generation of medieval studies interested in transcription from manuscript, are (1) do we have a diversity of digitised data to mitigate against what might be called "collections bias" described above, and (2) how much of this transcription, analysis and interpretation need to be done in order to understand fully enough what we can about any given corpus. In data science, it is recognized that predictive analytics can mean that we do not need to be concerned with getting all the data, but rather the right kinds of data. Will this be possible in medieval studies at present? Or will we need to do targeted digitization to be able to advance? It is also important to know when enough is enough. Incrementally adding more and more data will not necessarily produce better results, but sometimes results will simply plateau. The question we are asking ourselves at present is to what extent this will be the case with Latin Bibles.

## VII. USING OUR TRANSCRIPTIONS FOR ANALYTICAL PURPOSES

Our paper has addressed thus far the larger question of how medievalists have approached the issue of transcription when dealing with manuscripts and how those approaches are evolving with automation afforded by computer vision methods such as handwritten text recognition. A number of important issues about building corpora, creating balanced or even general models for medieval handwriting, even new challenges of error, bias and overfitting of machine learning models come to the fore. It is unlikely that we will be able to correct all the errors of these automated transcriptions with human labour, so we will need to find ways of looking for larger patterns while mitigating imprecision. In this section we will take a brief look at some kinds of research which simply could not be done without the kinds of transcriptions we have produced automatically, as well as new horizons of interpretation about medieval documents that open up with such transcriptions. Detailed descriptions about each of these projects will not be given, as it would take us beyond the scope of this article.

As was mentioned above, one of the simplest reasons that transcriptions of a manuscript are created is to facilitate searchability and keyword indexing. If we think about the ways we are able to search within modern texts such as pdfs or in online databases, "full text" searchability has become a basic expectation of researchers. Funded projects such as HIMANIS (HIstorical MANuscript Indexing for user-controlled Search, https://himanis.hypotheses.org) have leveraged computer vision techniques with voluminous archives such as chancery records to render these sources more friendly to the searching demands of researchers. A similar capability provided by the READ COOP's Transkribus Sites platform makes a first step in opening up handwritten archival documents to fuzzy searching.

Another important way that transcriptions from manuscript can enable analysis of medieval texts is in the domain of intertextuality. Transcriptions from textual traditions such as the *Roman de la Rose* could offer a window onto the *mouvance* of that text and the ways in which scribal culture modified it. In the context of our own work, we have found the equivalent of these intertextual elements in Latin Bibles where the text differs from the Vulgate. When sequences of strings do not match, cross checking with the Vetus Latina Database (Brepols) has revealed many examples where the language in our corpus of Paris Bibles echoes the pre-Vulgate text of the *Vetus Latina* (Table 5). These examples suggest an intertextuality based on orality and sermoning culture which influences in turn copying practices of bibles. Of course, a less conservative and more normalising approach to HTR would probably facilitate the computational identification of these variants more efficiently, but the principle of revealing intertextuality remains the same.

| Vulgate | LAD 2013.051 (Genesis) | VLB (Brepols) |
|---|---|---|
| et recordabor fœderis mei vobiscum, et cum omni anima vivente quæ carnem vegetat: et non erunt ultra aquæ diluvii ad delendum universam carnem. | Et recoꝛdaboꝛ feðeriſ mei **ꝙ pepiꝗi uobcum** 7 cum omi anima uiuente que carnem uegetat 7 ñ erunt ultra aque diluuii; að ðelenðam uniuſam carnē. | et recordabor foederis mei, **quod pepigi vobiscum** (PS-EUCH) |
| Dixitque ei angelus Domini: Revertere ad dominam tuam, et humiliare sub manu illius. | Dixitꝗ ei angĺs ðomini. Reůte að ðominam tuam 7 humilia-re ſub manu **iꝥius** | Dixitque angelus Domini Revertere ad dominam tuam, et humiliare sub manu **ipsius**. (PS-EUCH) |
| Vimque faciebant Lot vehementissime: jamque prope erat ut effringerent fores. | Uimꝗ faciebant loth ueheṁtiſ ſime iam ꝑpe erat ut **iſtrĩgent** foꝛes. | Vimque faciebant Lot vehementissime: jamque prope erat ut **infringerent** fores. (PS-EUCH) |
| duodecim duces generabit, et faciam illum in gentem magnam. | xii. ðuces genera-bit 7 faciã illũ **creſcere** in gentē magnam. | duodecim duces generabit et faciam illum **crescere** in gentem magnum (BREV GOTH) |
| Leva oculos tuos et vide a loco, in quo nunc es, ad aquilonem et meridiem, ad orientem et occidentem. | Leua oculos tuos **ĩ ðirectũ** 7 uiðe a loco ĩ quo nunc eſ a-quilonē 7 ṁiðiem að oꝛientem 7 occiðentem. | Gen. XXIII, ubi dicit Dominus ad Abraham: Leua oculos tuos **in directum** et uide a loco in quo nunc es, ad aquilonem et meridiem et orientem et occidentem. (Nicolas de Aquavilla) |

**Table 5**: Some examples of where departures from the Vulgate text in Genesis in the Louvre Abu Dhabi Bible (LAD 2013.051) match possible citations in the Vetus Latina Database (Brepols).

Another example of the possibility of searching specific words in the transcriptions made from manuscript comes from word counting. Using a concordancing tool, we can visually compare the frequencies of two versions of the Latin lemma *domin-* (meaning "Lord"), namely "ðn*" and "ðomin*". It can be seen that the variant ðn* is almost never detected in manuscript Smith-Lesouëf 19, whereas it is quite prevalent in the other three manuscripts compared here: BnF Latin 40 and BnF Latin 10421 (Figure 3). On the other hand, ðomin* is quite rare in LAD 2013.051 and very common in the New Testament in Smith-Lesouëf 19.

**CONCORDANCE DENSITY PLOTS (ðn* vs ðomin*)**

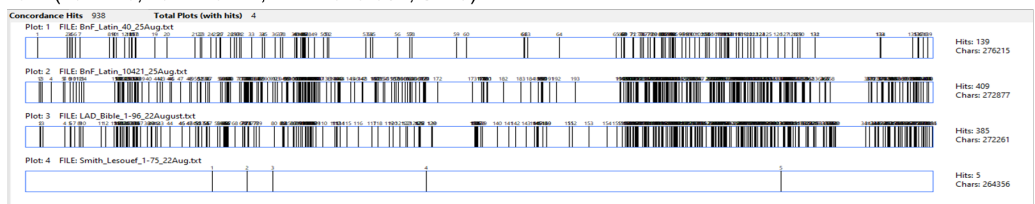

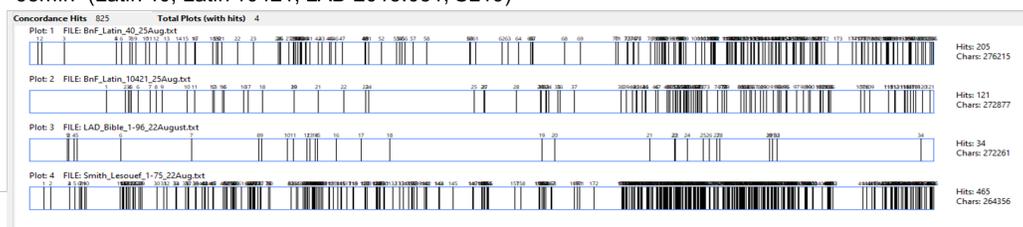

**Figure 3**: Concordance Density Plots made with AntConc for two strings representing the same lemma (ðn* and ðomin*) for four manuscripts of Latin Bibles (BnF Smith-Lesouëf 19, BnF Latin 40, BnF Latin 10421, LAD 2013.051). Visualised with AntConc.

These differences allow us to look at variance in individual groups of manuscripts for a single feature indicative of a scribal profile, but also contribute to computational methods we can use across many manuscripts and features to perform classification experiments. In this respect, the title of this article "Transcription of Medieval Manuscripts for Machine Learning" not only points to modes of transcription which allow for HTR transcriptions to be produced, but those to the study of those transcriptions using machine learning to identify larger scale patterns. Of course, not all corpora of interest to the medievalist might qualify in size or quality for this kind of analysis, but it does open the door to forms of predictive analysis for some.

For example, in the case of the manuscript Cambridge, Corpus Christi College 49, known to exhibit three different hands, we are able to use the technique known as rolling stylometry to predict computationally the identity of those hands, which we have previously confirmed with our human eye. Significantly, using this method of sequential analysis, we are able to predict the identity of the copyist at any given point of the manuscript with a quite small sample of language, down to less than 1500 words. As Kestemont (2015) wrote: "superficial textual variations also present important scholarly opportunities, for instance for the identification of scribes or the dialectological analysis of texts." Several scholars have also discussed the issues caused by data loss linked to the normalisation of transcriptions and the discarding variations that are considered random or trivial (Driscoll 2006; Kytö et al. 2011; Kopaczyk 2011; Rogos 2011; 2012; Stutzman 2014; Lass 2004). This criticism is supported by the fact that there are numerous approaches which have demonstrated the value of scribal and other accidental variation. The method of predictive analysis of the hand of a known scribe demonstrated here for the case of one manuscript confirms what we know from close visual analysis, but also has some remarkable possibilities when working with other transcriptions, in combination with material philological evidence for example, for gaining a better understanding about how such manuscripts were made and transcribed.

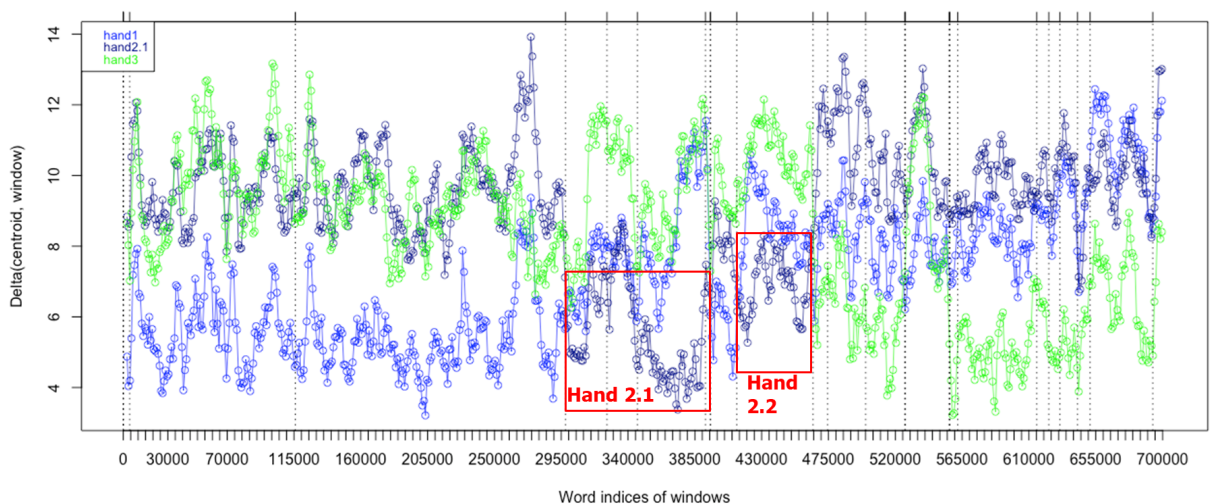

**Figure 4**: A graphic output of an experiment using the rolling delta method for predictive classification of the scribal hands of a Latin Bible, Cambridge, Corpus Christi College 49. Visualised in R with the Stylo package.

One last example is a very distant analysis. Using the method known as TF-IDF (or Term Frequency - Inverse Document Frequency), we can look at a number of non-normalised transcriptions of segments of manuscripts from the Paris Bible tradition in order to predict dating or localization based on what we known from legacy metadata. TF-IDF, as a classification method, helps us to approximate common words in each transcription, while balancing its importance in other documents. Clusters of transcriptions which share similar common words (and necessarily common abbreviations and words spelled with special letter-forms) cluster together, but not with the rest of the corpus.

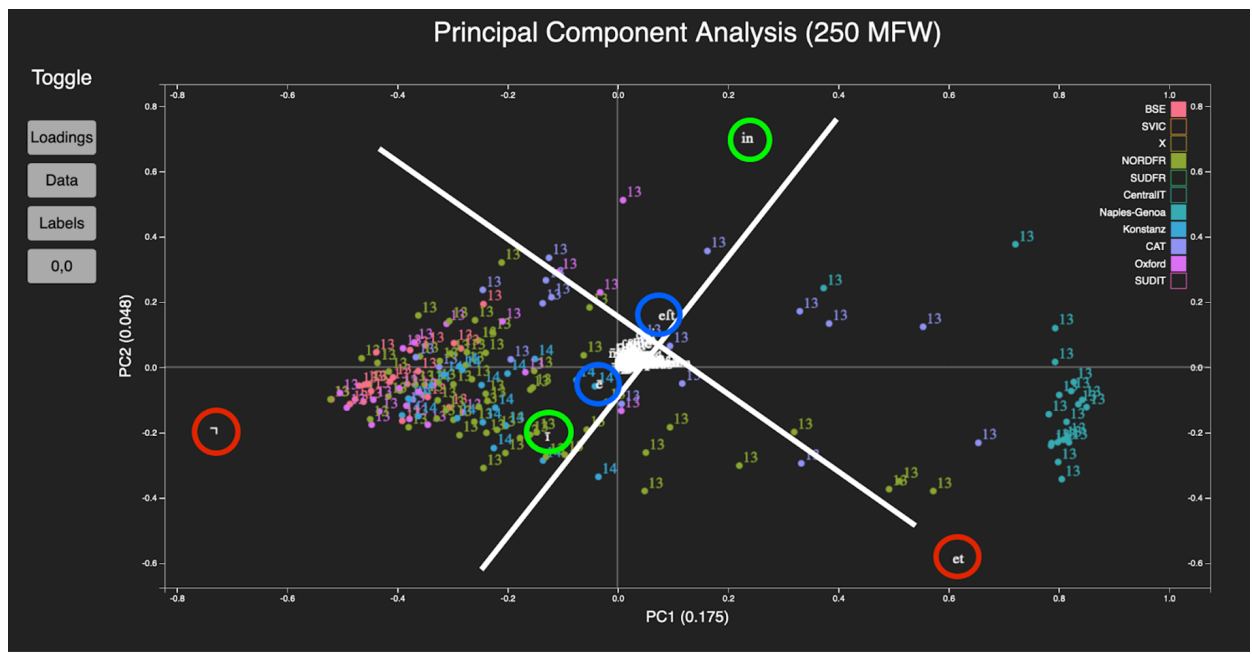

**Figure 5**: Principal Component Analysis (PCA) with loadings resulting from TF-IDF analysis of 24 transcriptions of Latin Bibles from many different regions, carried out using scikit. Code adapted from Paul Vierthaler.

Using a custom word dictionary, what we have found using this method is that there are certain high frequency words with specific spellings (in / ī and eſt / ē from top right to bottom left) and (et / ⁊) from top left to bottom right which are indicative of regions known from the localization of manuscripts. This tends to suggest that English scribes are more likely to use the Tirolian ampersand, whereas Italian or some French scribes preferred the two letter "et". A similar distinction is made between Catalonian and English scribes' preference for the preposition "in" and French scribes "ī". Such results might seem basic, but they suggest new ways we can use non-normalised transcription as an additional layer of data for scribal attribution, dating and localization. This analysis is suggestive in as much as it could help us build a handlist of high-level features distinctive of specific regions.

Of course, the two perspectives we have explored here, diplomatic and normative transcription, should not be stated as opposite views but rather as different layers or steps in scholarly workflows. We argue for rethinking transcription as a process of working with multiple layers of data that imagines different forms of analysis and different research questions. From diplomatic forms, mixing graphic and the graphetic levels, one can still expand and normalise for a scholarly edition in a second step. On the other hand, transcribing while normalising, one cannot go backward to study shifts in scribal hand; data has simply been lost. Normalisation can be rethought as a second step, and could even be designed an automated task. Two caveats for such a process: there could also be ambiguities in the case of abbreviated words and there are inevitable modifications to the ways we think of transcribing and editing using HTR. Such work in the age of automation requires us to redesign the way we work.

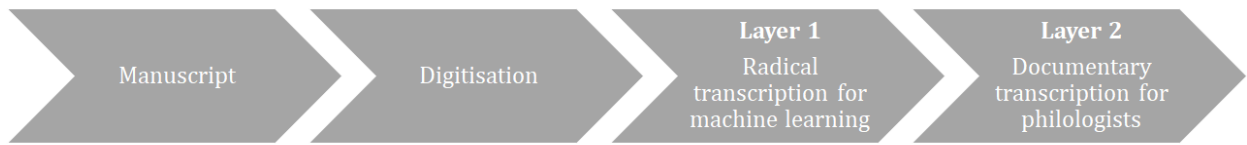

**Figure 6**: A workflow for moving from manuscript to multiple layers of transcription.

## VIII. A CONTINUUM OF BIAS

In our paper, we have argued for the importance of an explicit approach to transcription norms which encode evidence about our research documents and link them to specific research questions, with an emphasis on reproducibility and plain text environments. We also have explained how, along with principles of transcription, the material quality of digitised collections and their basic availability via download or IIIF impacts how we are able to create a corpus of transcriptions for computational study. In closing we have presented some examples of computational analysis of the non-normalised transcription methods we propose.

In the not-so-distant future, we predict that mainstream medieval studies will be using much more text which has been transcribed directly from manuscript, and the use of semi-automated methods for text extraction will be more widespread than they are now. As we have argued in our paper, automated methods make possible the inclusion of micro-features in transcription schemes for capturing different kinds of data about scribal practice in manuscripts. Conversely, HTR models can also facilitate certain normalisation gestures in the process of manuscript transcription. Whereas this categorization might seem to cast transcription in terms of two extremes–one conservative and one normalising–a more likely future is one in which there are many different customised methods for accessing text in the pursuit of specific research questions. Even more likely is that researchers of the future will encounter versions of texts across the spectrum of normalisation, and they will want to be able to work with all of them, analogous to the way that edition-centred scholarship traditionally used critical editions of texts which did not share the same degree of editorial intervention alongside each other.

It is likely that future text processing methods will emerge to handle these discrepancies in and between texts. In the era of growing popularity of HTR and of data sharing, there is, in our opinion, a new responsibility for the medievalist to participate in thoughtful and explicit text creation. It is unlikely that we will eliminate bias in HTR models, since after all, the transcription norms that we employ based on specific textual traditions are themselves forms of bias. Furthermore, datasets combines very different domains and kinds of textual artefacts. Whereas the process of

Journal of Data Mining and Digital Humanities                                    http://jdmdh.episciences.org
ISSN 2416-5999, an open-access journal

remediating medieval texts has been described in terms of the so-called "unedition" (Dark Archives, 2021), one of the pieces of the traditional edition we strongly believe should be preserved moving forward is the editorial statement. Since the relationship between transcription, manuscript work and technology has changed—and will continue to change—our scholarly practices, let us call this new version a "transcription statement." In such a statement, the need to outline one's theoretical and practical principles for transcription norms, the anticipated research questions that a particular layer of data might uncover, samples of training and output data, as well as the principles of model training and correction ought to be provided. The HTR United project has been created to bring some of this information together in a metadata catalogue, however, more work can be done in this domain. In our paper we have attempted to model these kinds of observations using details from our own experiments with transcription. Akin to the calls for multiple forms of transparency about the process and context of data creation (D'Ignazio and Klein, 2020), we imagine such "transcription statements" providing a point of access for understanding what design decisions were made in the creation of the model, unpacking in detail what kinds of information have been privileged and by whom, what kinds of bias are embedded in libraries, in codices, in particular their digitised versions, as well as how those various levels of bias impede our field's deeper understanding of our new objects of study on the way to a future of digital manuscript studies.